\documentclass[a4paper,12pt]{article}
\usepackage[T1]{fontenc}
\usepackage[utf8]{inputenc}
\usepackage[spanish,english]{babel}
\usepackage{authblk}
\usepackage{csquotes}
\usepackage{xpatch}
\usepackage{hyperref}
\hypersetup{hidelinks}
\usepackage[hypcap=false]{caption}
\usepackage[font=small,labelfont=bf]{caption}
\usepackage{amsmath}
\usepackage{amssymb}
\usepackage{amsthm}
\usepackage{mathtools}
\usepackage{mathrsfs}
\usepackage{braket}
\usepackage{bm}
\usepackage{graphicx}
\usepackage{booktabs}
\usepackage{siunitx}
\usepackage{feynmf}

\newcommand{\mail}[1]{\href{mailto:#1}{\texttt{#1}}}
\newcommand{\keywords}[1]{\textbf{Keywords:} #1}

\title{\textbf{Quark model description of $\psi(4260)$}}
\author{R. Bruschini\thanks{\mail{roberto.bruschini@ific.uv.es}}}
\author{P. González\thanks{\mail{pedro.gonzalez@uv.es}}}
\affil{\foreignlanguage{spanish}{Departamento de Física Teórica-IFIC \\
	Universidad de Valencia-CSIC \\
	E-46100 Burjassot(Valencia)}, Spain}
\date{}

\begin{document}

\maketitle

\begin{abstract}
From lattice indications we follow a Born-Oppenheimer approximation to build a quark-antiquark static potential for $J^{PC}=1^{--}$ charmonium states below their first S- wave meson-meson threshold. We show that a good description of the mass and decay properties of the experimentally well established $\psi(4260)$ resonance is feasible.
\end{abstract}

\begin{center}
\end{center}

\section{Introduction\label{SI}}

The explanation of experimentally discovered charmonium states, that do not
fit well in conventional quark model descriptions of heavy quarkonia as for
instance the ones provided by the Cornell \cite{Eic80,Eic04} or the
Godfrey-Isgur \cite{GI85} models, is nowadays a theoretical challenge.

Regarding unconventional isospin $0$ states ($\chi_{c1}\left(  3872\right)  $,
$\psi(4260)$ $...)$, see \cite{PDG18}, the presence of close open flavor
(charm) meson-meson thresholds may be playing an important role. As a matter
of fact explanations involving the presence of meson-meson components in the
form of either molecules, or tetraquarks implicitly involving several
molecular configurations, or complementary configurations to the heavy
quark-antiquark ones have been developed (for recent bibliographic reviews see
\cite{Ch16}, \cite{Leb17}, \cite{Guo17}, \cite{Esp17} and references therein;
for a more general heavy quarkonia review see \cite{Bra11}).

\bigskip

One possible alternative explanation may come from the consideration of a
\textquotedblleft beyond the conventional\textquotedblright\ quark model
description, where the meson-meson degrees of freedom as well as the gluon
ones are integrated out through an effective heavy quark-antiquark potential.
A specific form for this potential can be proposed from lattice calculations
\cite{Bal05} for the energy of two static color sources (quark $Q$ and
antiquark $\overline{Q}$) when mixing of the $Q\overline{Q}$ configuration
with an open flavor meson-meson one is taken into account. By using a
Born-Oppenheimer approximation the resulting $Q\overline{Q}$ static potential
below the meson-meson threshold exhibits screening starting at a certain
energy below the threshold and saturating (becoming flat) at the threshold mass.

The screening energy interval is shorter for $Q\overline{Q}$ configurations
($Q=b$ or $c$) involving only mesons with very small widths ($B,B^{\ast}$ or
$D,D^{\ast}$). From lattice results the starting screening energy in this case
may be estimated to be about $30$ MeV below the threshold mass. In a first
approach one may tentatively take the simplifying assumption that screening
takes place just at the threshold mass (zero screening energy interval
approach). This idea has been implemented and extended through the so called
Generalized Screened Potential Model for the description of $0^{+}\left(
J^{++}\right)  $ charmonium $\left(  J=0,1,2\right)  $ \cite{Gon15,Gon17} as
well as bottomonium states \cite{Gon14}.

When applied to $0^{-}\left(  1^{--}\right)  $ charmonium this approach fails
even in the low energy spectral region below the first $S-$ wave meson-meson
threshold that we shall call henceforth $D\overline{D_{1}}$ (involving
$D\overline{D_{1}}\left(  2420\right)  $ and $D\overline{D_{1}}\left(
2430\right)  $) since the unconventional $\psi(4260)$ can not be sensibly
assigned to any state from the potential. This failure may have to do with the
need of implementing a non zero screening energy interval in this case due to
the significant threshold width.

\bigskip

In this article we try to go a step further in the construction of the
potential for $0^{-}\left(  1^{--}\right)  $ charmonium by implementing a non
zero screening energy interval. We shall show that a reasonable description of
$0^{-}\left(  1^{--}\right)  $ states lying below the $D\overline{D_{1}}$
threshold including $\psi(4260)$ may be attained. The contents of the article
are organized as follows. In Section \ref{SII} a brief review of the (zero
screening energy interval approach) potential for $0^{+}\left(  1^{++}\right)
$ states below their first $S-$ wave meson-meson threshold is presented. In
Section \ref{SIII} we implement the potential for $0^{-}\left(  1^{--}\right)
$ states. From it we calculate the spectrum below their first $S-$ wave
meson-meson threshold. In Section \ref{SIV} we concentrate on the study of
$\psi(4260),$ the only well established unconventional state in this spectral
region. We calculate its decay properties and compare them to existing data.
Finally in Section \ref{SV} our main results and conclusions are summarized.

\section{$c\overline{c}$ Potential for $0^{+}\left(  1^{++}\right)  $
states\label{SII}}

In order to construct a $Q\overline{Q}$ static potential implicitly
incorporating the effect of meson-meson components we shall start from
(unquenched) lattice results \cite{Bal05} for the energy of two static color
sources ($Q$ and $\overline{Q}$) when mixing of the $Q\overline{Q}$
configuration with an open flavor meson-meson one is taken into consideration.
As a consequence of the presence of this meson-meson configuration the
$Q\overline{Q}$ static energy changes its radial dependence on the
$Q-\overline{Q}$ distance. Following a Born-Oppenheimer interpretation we
shall identify the $Q\overline{Q}$ static energy with the $Q\overline{Q}$
static potential (for a review of Born-Oppenheimer potentials see
\cite{Braa14}).

\bigskip

\begin{figure}
[ptb]
\begin{center}
\includegraphics[
height=2.3808in,
width=3.659in
]%
{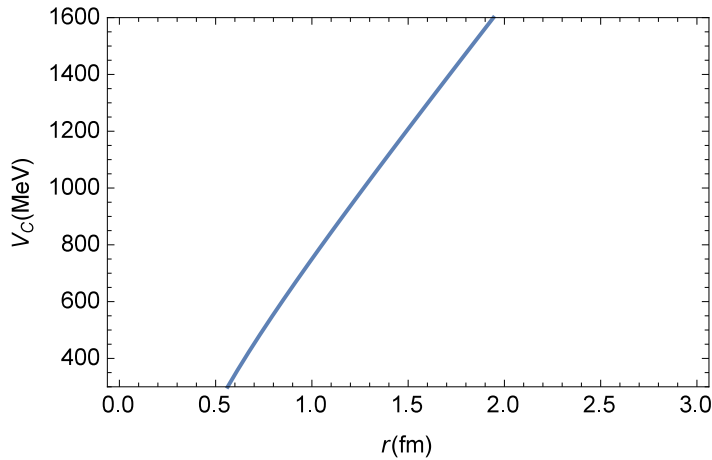}%
\caption{Representation of the static potential $V_{C}(r)$ with $\sigma=850$
MeV/fm and $\zeta=100$ MeV.fm.}%
\label{Fig1}%
\end{center}
\end{figure}

For the sake of clarity let us go step by step. First let us only consider a
$Q\overline{Q}$ configuration. The dependence of the $Q\overline{Q}$ static
energy on the $Q-\overline{Q}$ distance has been derived in (quenched) lattice
QCD \cite{Bal01}. By identifying this energy dependence with the (quenched)
$Q\overline{Q}$ static potential one gets a Cornell like form%
\begin{equation}
V_{C}\left(  r\right)  =\sigma r-\frac{\zeta}{r} \label{Cornell}%
\end{equation}
where $r$ is the $Q-\overline{Q}$ distance and the parameters $\sigma$ and
$\zeta$ stand for the string tension and the chromoelectric coulomb strength
respectively. This potential is drawn in Fig. \ref{Fig1} where the values of
the parameters
\begin{equation}%
\begin{array}
[c]{c}%
\sigma=850\text{ MeV/fm}\\
\zeta=100\text{ MeV.fm}\\
m_{c}=1348.6\text{ MeV}\\
m_{b}=4793\text{ MeV}%
\end{array}
\end{equation}
have been chosen to get a reasonable fit of the low lying $0^{+}\left(
J^{++}\right)  $ charmonium and bottomonium spectra \cite{Gon15,Gon14}.%

\bigskip

Let us now consider a $Q\overline{Q}$ configuration with quantum numbers
$I^{G}\left(  J^{PC}\right)  $, for example $0^{+}\left(  1^{++}\right)  $
$c\overline{c}$, plus a meson-meson configuration. It is important to realize
that the first open flavor meson-meson configuration with these quantum
numbers that may contribute to the static potential is $D^{0}\overline
{D^{\ast0}}$ (from now on it is always understood that the sum of the charge
conjugate meson-meson configuration is implicit). This is so, despite the fact
that $D^{0}\overline{D}^{0}$ has a lower energy threshold, because the two
mesons have to be in an $S-$ wave channel for the $c$ quark in one meson and
the $\overline{c}$ antiquark in the other meson to remain static as required
(this is only strictly true in the infinite $c$ mass limit, $m_{c}%
\rightarrow\infty,$ but it can be taken as a good approximation). (Actually,
$D^{0}\overline{D}^{0}$ is the first threshold contributing to the
$0^{+}\left(  0^{++}\right)  $ static potential.)

\bigskip

From lattice results obtained when a $Q\overline{Q}$ and a meson-meson
configurations are considered \cite{Bal05} we expect $c\overline{c}$ and
$D\overline{D^{\ast}}$ mixing (for simplicity we assume the same mass for the
different isospin components and call the threshold $D\overline{D^{\ast}}$).
This makes the formal dependence of the $c\overline{c}$ static energy on the
$c-\overline{c}$ distance to be different close below and above the threshold
mass. In particular, this dependence starts to differ from the Cornell like
form when approaching the meson-meson threshold from below becoming flat at
the threshold mass. If this change takes place in a small energy region then
the identification of this energy dependence with a (unquenched) $0^{+}\left(
1^{++}\right)  $ $c\overline{c}$ static potential gives rise to the
approximate form%
\begin{equation}
V_{\left[  0,\text{ }m_{D\overline{D^{\ast}}}\right]  }(r)=\left\{
\begin{array}
[c]{c}%
\sigma r-\frac{\zeta}{r}\text{
\ \ \ \ \ \ \ \ \ \ \ \ \ \ \ \ \ \ \ \ \ \ \ \ \ \ \ }r\leq r_{D\overline
{D^{\ast}}}\\
\\
m_{D\overline{D^{\ast}}}-m_{c}-m_{\overline{c}}\text{
\ \ \ \ \ \ \ \ \ \ \ \ \ \ }r\geq r_{D\overline{D^{\ast}}}%
\end{array}
\right.  \label{Pot1++}%
\end{equation}
where the bracket subindex $\left[  0,m_{D\overline{D^{\ast}}}\right]  $
indicates that this potential is only valid up to the threshold mass
$m_{D\overline{D^{\ast}}}=m_{D}+m_{\overline{D^{\ast}}},$ and the crossing
radii $r_{D\overline{D^{\ast}}}$ is defined by the continuity of the potential
at the threshold as%
\begin{equation}
\sigma r_{D\overline{D^{\ast}}}-\frac{\zeta}{r_{D\overline{D^{\ast}}}%
}=m_{D\overline{D^{\ast}}}-m_{c}-m_{\overline{c}} \label{rDDbar}%
\end{equation}

\begin{figure}
[ptb]
\begin{center}
\includegraphics[
height=2.4396in,
width=3.659in
]%
{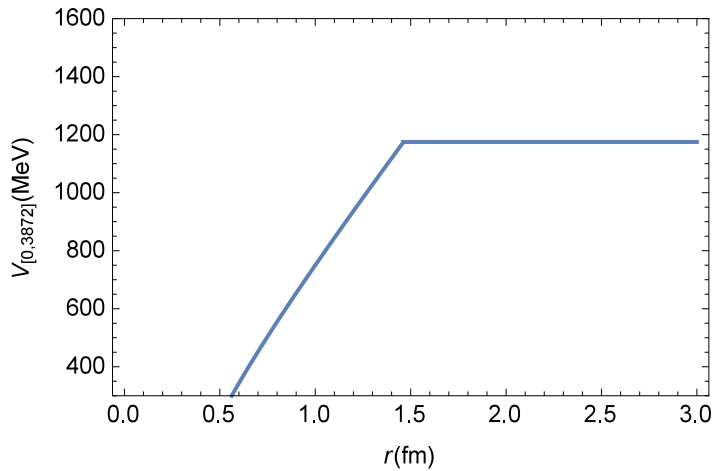}%
\caption{Representation of the $0^{+}(1^{++})$ $c\overline{c}$ static
potential $V_{\left[  0,\text{ }m_{D\overline{D^{\ast}}}\right]  }(r)$ with
$m_{c}=1348.6$ MeV, $\sigma=850$ MeV/fm, $\zeta=100$ MeV.fm and $m_{D\overline
{D^{\ast}}}=3872$ MeV.}%
\label{Fig2}%
\end{center}
\end{figure}

This potential, corresponding to a zero screening energy interval approach,
has been drawn in Fig. \ref{Fig2} for the same values of the parameters
previously used for $V_{C}\left(  r\right)  .$ As for the threshold mass we
use the value $m_{D\overline{D^{\ast}}}=3872$ MeV obtained from the
experimental masses of $D$ and $\overline{D^{\ast}}$ \cite{PDG18}.%

\bigskip

The physical mechanism underlying this potential has to do with the creation
of $q\overline{q}$ pairs, where $q$ stands for a light quark $\left(
q=u,d,s\right)  ,$ and the later combination of $q\left(  \overline{q}\right)
$ with $\overline{c}\left(  c\right)  $ giving rise to a total screening of
the $c$ and $\overline{c}$ color charges at the threshold mass (string
breaking) since the formed mesons $D$ and $\overline{D^{\ast}}$ are color singlets.

\bigskip

It is worth to remark that whereas $V_{C}\left(  r\right)  $ is defined in the
whole spectral energy region the potential $V_{\left[  0,\text{ }%
m_{D\overline{D^{\ast}}}\right]  }(r)$ can only be applied to calculate
$0^{+}\left(  1^{++}\right)  $ charmonium states with mass below the
$D\overline{D^{\ast}}$ threshold mass. Therefore it is a confining potential.
For higher energies the form of the $c\overline{c}$ static potential is
different (one possible choice has been used to build the Generalized Screened
Potential Model \cite{Gon14}).

To get the low lying $0^{+}\left(  1^{++}\right)  $ charmonium spectrum up to
$m_{D\overline{D^{\ast}}}$ we solve the Schr\"{o}dinger equation for
$V_{\left[  0,\text{ }m_{D\overline{D^{\ast}}}\right]  }(r)$. The results
obtained are listed in Table~\ref{Tab1}. Notice that we assign our calculated
states to spin triplet ones; the reason is that our potential is spin
independent and we know that spin-spin corrections to the mass are bigger (by
a factor 3) for spin singlet than for spin triplet states. \begin{table}[ptb]%
\begin{tabular}
[c]{cccccc}%
$J^{PC}$ & $%
\begin{array}
[c]{c}%
\text{States}\\
np_{\left[  0,\text{ }m_{D\overline{D^{\ast}}}\right]  }%
\end{array}
$ & $%
\begin{array}
[c]{c}%
m_{\left[  0,\text{ }m_{D\overline{D^{\ast}}}\right]  }\\
\text{MeV}%
\end{array}
$ & $%
\begin{array}
[c]{c}%
m_{PDG}\\
\text{MeV}%
\end{array}
$ & $%
\begin{array}
[c]{c}%
m_{Cor}\\
\text{MeV}%
\end{array}
$ & $%
\begin{array}
[c]{c}%
V_{C}\left(  r\right)  \text{ States }\\
np
\end{array}
$\\\hline
$1^{++}$ &  &  &  &  & \\
& $1p_{\left[  0,\text{ }m_{D\overline{D^{\ast}}}\right]  }$ & $3454.8$ &
$3510.66\pm0.07$ & $3456.2$ & $1p$\\
&  &  &  &  & \\
& $2p_{\left[  0,\text{ }m_{D\overline{D^{\ast}}}\right]  }$ & $3871.7$ &
$3871.69\pm0.17$ &  &
\end{tabular}
\caption{Calculated $0^{+}\left(  1^{++}\right)  $ charmonium masses
$m_{\left[  0,\text{ }m_{D\overline{D^{\ast}}}\right]  }$ from $V_{\left[
0,\text{ }m_{D\overline{D^{\ast}}}\right]  }(r)$ with $\sigma=850$ MeV/fm,
$\zeta=100$ MeV.fm, $m_{c}=1348.6$ MeV and $m_{D\overline{D^{\ast}}}=3872$
MeV. The spectral notation $np_{\left[  0,\text{ }m_{D\overline{D^{\ast}}%
}\right]  },$ where $n$ $\left(  p\right)  $ indicates the principal (orbital
angular momentum) quantum number has been used for the states. Masses for
experimental resonances, $m_{PDG},$ have been taken from \cite{PDG18}. Masses
$m_{Cor}$ from $V_{C}(r),$ up to $m_{D\overline{D^{\ast}}}$, with the same
values for $\sigma,$ $\zeta$ and $m_{c}$ are also shown for comparison. }%
\label{Tab1}%
\end{table}

\begin{figure}
[ptb]
\begin{center}
\includegraphics[
height=2.3964in,
width=3.659in
]%
{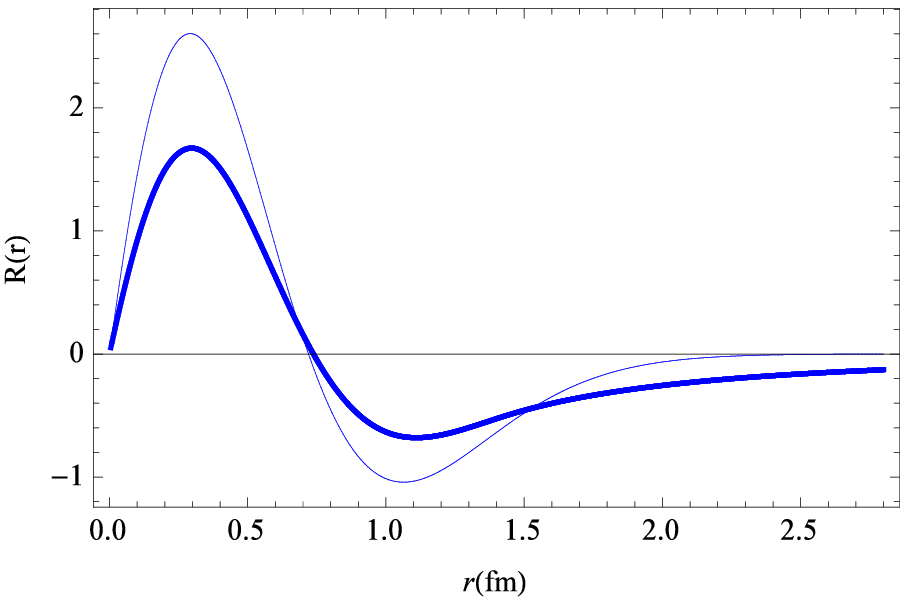}%
\caption{Radial wave functions R(\textit{r}) (in units $fm^{-\frac{3}{2}}$)
for the $1^{++}\left(  2p_{\left[  0,m_{D\overline{D^{\ast}}}\right]
}\right)  $ state (thick line) and the $1^{++}\left(  2p\right)  $ state (thin
line).}%
\label{Fig3}%
\end{center}
\end{figure}

Let us realize that there is no difference between the $1p_{\left[  0,\text{
}m_{D\overline{D^{\ast}}}\right]  }$ and the conventional $1p$ state since
quite below threshold there is no difference between using $V_{\left[
0,\text{ }m_{D\overline{D^{\ast}}}\right]  }(r)$ and $V_{C}(r)$. On the
contrary there is a big difference between the $2p_{\left[  0,\text{
}m_{D\overline{D^{\ast}}}\right]  }$ state lying below threshold and the
conventional $2p$ state with mass above it $\left(  m_{Cor}\left(  2p\right)
=3911\text{ MeV}\right)  $. A justified assignment of the $2p_{\left[
0,\text{ }m_{D\overline{D^{\ast}}}\right]  }$ state to $\chi_{c1}\left(
3872\right)  $ has been done elsewhere \cite{Gon15}. Here we just plot in Fig.
\ref{Fig3} the $2p_{\left[  0,\text{ }m_{D\overline{D^{\ast}}}\right]  }$
radial wave function as compared to the $2p$ radial wave function to make
clear the difference between them. We observe that as a consequence of the
color screening in the $2p_{\left[  0,\text{ }m_{D\overline{D^{\ast}}}\right]
}$ state there is a flux of probability from the origin outwards as compared
to the non screened case. This will be important for the numerical evaluation
of the width for the electromagnetic transition between $\psi(4260)$ and
$\chi_{c1}\left(  3872\right)  $.%

\section{$c\overline{c}$ Potential for $0^{-}\left(  1^{--}\right)  $ states
\label{SIII}}

When considering the $0^{-}\left(  1^{--}\right)  $ case the simple
prescription of a zero screening energy interval adopted for the construction
of the potential for $0^{+}\left(  1^{++}\right)  $ states has to be refined
if we want to accommodate the existing data. As said before the same approach
can not give any state to be reasonably assigned to $\psi(4260).$

For this refinement let us remind that the first $S-$ wave meson-meson
threshold for $0^{-}\left(  1^{--}\right)  $ states, $D\overline{D_{1}},$ with
a threshold mass $m_{D\overline{D_{1}}}\simeq4287$ MeV, corresponds to
$D^{0}\overline{D_{1}}^{0}(2420)$ where $\overline{D_{1}}^{0}(2420)$ has a
width of about $30$ MeV and to $D^{0}\overline{D_{1}}^{0}(2430)$ where
$\overline{D_{1}}^{0}(2430)$ has a larger but quite uncertain width
($384_{-75}^{+107}\pm74$ MeV). As said before the threshold effect on the
static potential comes from the coupling of $Q\overline{Q}$ to light
$q\overline{q}$ pairs out of the vacuum giving rise to the meson-meson
threshold components. It turns out that in the limit $m_{Q}\rightarrow\infty$
the strong interaction has Heavy Quark Spin Symmetry (HQSS) and this prevents
the formation of $D^{0}\overline{D_{1}}^{0}(2420)$ from a $Q\overline
{Q}\left(  1^{--}\right)  $ and a $q\overline{q}\left(  0^{++}\right)  $
\cite{LiVo13}. Therefore in this limit the only meson-meson component to be
taken into account in the construction of the potential should be
$D^{0}\overline{D_{1}}^{0}(2430).$ One should consider though that HQSS
breaking is expected given the real (non infinite) mass of the charm quark, as
detailed in reference \cite{Wan14}. Hence we shall consider $D\overline{D_{1}%
}$ as an effective threshold that may be also incorporating the possible
effect of $D^{0}\overline{D_{1}}^{0}(2420)$. It is physically reasonable to
assume that due to the non negligible widths of $\overline{D_{1}}^{0}(2430)$
and $\overline{D_{1}}^{0}(2420)$ the starting screening energy in the
$0^{-}\left(  1^{--}\right)  $ case is lying quite below threshold as compared
to the $0^{+}\left(  J^{++}\right)  $ case where the threshold widths are
negligible. If we remind that for the $0^{+}\left(  0^{++}\right)  $ case
lattice calculations give a starting screening energy of about $30$ MeV below
threshold then, from the $\overline{D_{1}}^{0}(2420)$ width $\left(
\simeq30\text{ MeV)}\right)  ,$ we may reasonably expect for the $0^{-}\left(
1^{--}\right)  $ case the starting screening energy to be at least $60$ MeV
below threshold. To be more specific let us call the starting screening energy
$E_{s}\equiv m_{D^{0}\overline{D_{1}}^{0}}-m_{c}-m_{\overline{c}}-\Delta$
where $\Delta$ indicates its distance to the threshold. Then according to our
expectation $\Delta\geq60$ MeV. On the other hand we expect $\Delta$ to be
limited by a value $30$ MeV bigger than the value of the $\overline{D_{1}}%
^{0}(2430)$ width. This determines the expected interval of possible values
for $\Delta.$ Unfortunately the uncertainty in the knowledge of the
$\overline{D_{1}}^{0}(2430)$ width does not permit to fix precisely the upper
bound for $\Delta.$ Instead we shall use in what follows the scarce
$\psi(4260)$ data to try to fix it as much as possible. This will allow us to
conclude that values of $\Delta$ within the interval $\left[  60,120\right]  $
MeV may give quantitative account of the observed properties of $\psi(4260),$
see below.

\bigskip

The static potential will start to differ from $V_{C}(r)$ at $E_{s}$. To take
this into account in a simple manner we shall assume that at $E_{s}$ the
potential reduces its slope (the one from $V_{C}(r)$) to a constant value$\ s$
which is maintained up to the threshold mass where it becomes $0$. This should
be considered as an average approximation to the gradual decreasing of the
slope that it is expected to really take place.

Specifically the proposed potential for $0^{-}\left(  1^{--}\right)  $ states
reads (again we shall assume isospin symmetry)
\begin{equation}
V_{\left[  0,m_{D\overline{D_{1}}}\right]  }(r)=\left\{
\begin{array}
[c]{c}%
\sigma r-\frac{\zeta}{r}%
\text{\ \ \ \ \ \ \ \ \ \ \ \ \ \ \ \ \ \ \ \ \ \ \ \ \ \ \ \ \ \ \ \ \ \ \ \ \ \ \ \ \ \ \ \ \ \ \ \ \ \ \ \ \ \ \ \ \ \ \ \ \ }%
r\leq r_{\Delta}\\
\\
\left(  m_{D\overline{D_{1}}}-m_{c}-m_{\overline{c}}-\Delta\right)  +s\left(
r-r_{\Delta}\right)  \text{\ \ \ \ \ \ \ \ \ \ \ \ \ }r_{\Delta}\leq
r\leq\left(  r_{\times}\right)  _{D\overline{D_{1}}}\\
\\
m_{D\overline{D_{1}}}-m_{c}-m_{\overline{c}}\text{
\ \ \ \ \ \ \ \ \ \ \ \ \ \ \ \ \ \ \ \ \ \ \ \ \ \ \ \ \ \ \ \ \ \ \ \ \ \ \ }%
r\geq\left(  r_{\times}\right)  _{D\overline{D_{1}}}%
\end{array}
\right.  \label{Pot1--}%
\end{equation}
where $r_{\Delta}$ and $\left(  r_{\times}\right)  _{D\overline{D_{1}}}$ are
defined by the continuity of the potential as%
\begin{equation}
\sigma r_{\Delta}-\frac{\zeta}{r_{\Delta}}=m_{D\overline{D_{1}}}%
-m_{c}-m_{\overline{c}}-\Delta\label{rDelta}%
\end{equation}%
\begin{equation}
-\Delta+s\left(  \left(  r_{\times}\right)  _{D\overline{D_{1}}}-r_{\Delta
}\right)  =0 \label{slope}%
\end{equation}

\bigskip

Regarding the value of the slope $s,$ we shall fix it by requiring that a
bound state close below threshold appears as experimentally required by the
presence of the unconventional $\psi(4260)$ resonance. It turns out that $s$
and $\Delta$ are correlated in the sense that an increasing of $\Delta$ can be
compensated by an increasing of $s$ to get the same mass for the bound state,
as can be checked in Table~\ref{Tab1b}. This mitigates the lack of a clear
connection between the chosen value of $\Delta$ and the real threshold widths.

\begin{center}
\begin{table}[ptb]%
\begin{tabular}
[c]{cccc}%
$%
\begin{array}
[c]{c}%
\Delta\\
\text{MeV}%
\end{array}
$ & $%
\begin{array}
[c]{c}%
s\\
\text{MeV/fm}%
\end{array}
$ & $%
\begin{array}
[c]{c}%
m_{4s_{\left[  0,m_{D\overline{D_{1}}}\right]  }}\\
\text{MeV}%
\end{array}
$ & $%
\begin{array}
[c]{c}%
\left\langle r^{2}\right\rangle ^{\frac{1}{2}}\\
\text{fm}%
\end{array}
$\\\hline
&  &  & \\
$0$ & $\times$ &  & \\
&  &  & \\
$30$ & $\times$ &  & \\
&  &  & \\
$60$ & $13$ & $4261.5$ & $3.7$\\
&  &  & \\
$120$ & $64.2$ & $4261.5$ & $2.8$\\
&  &  & \\
$180$ & $135$ & $4261.5$ & $2.5$%
\end{tabular}
\caption{Correlated $\Delta$\ and $s$\ values giving rise to the same mass for
the $4s_{\left[  0,m_{D\overline{D_{1}}}\right]  }$ state from $V_{\left[
0,m_{D\overline{D_{1}}}\right]  }(r)$ with $\sigma=850$ MeV/fm, $\zeta=100$
MeV.fm, $m_{c}=1348.6$ MeV and $m_{D\overline{D_{1}}}=4287$ MeV. Calculated
root mean square values $\left\langle r^{2}\right\rangle ^{\frac{1}{2}}$ are
also listed. The $\times$ sign indicates that no value of $s$ can be found to
get the required bound state.}%
\label{Tab1b}%
\end{table}
\end{center}

It should be emphasized that for $\Delta=0$ no bound state that could be
assigned to $\psi(4260)$ can be generated. This is the quantitative
translation of our previous comment about the need of going beyond the zero
screening energy interval approach for $0^{-}\left(  1^{--}\right)  $ states.
For $\Delta\leq30$ MeV the only possibility to generate bound states close
below threshold is by choosing such a small value of $s$ that an unphysical
proliferation of bound states occur. Only for $\Delta\gtrsim60$ MeV a well
defined bound state with the required mass of $4260$ MeV appears. Regarding
other states than $\psi(4260),$ with masses below $4200$ MeV, the different
$\left(  \Delta,s\right)  $ pairs considered produce a rather small change in
the mass of the high lying ones, of $10$ MeV at most , giving rise to quite
the same spectral description. (Notice that the higher the $\Delta$ value the
bigger the change in the masses, what indicates that $\Delta$ cannot be much
larger than $180$ MeV for the same spectrum to be maintained.)

\bigskip

\begin{figure}
[ptb]
\begin{center}
\includegraphics[
height=2.4111in,
width=3.659in
]%
{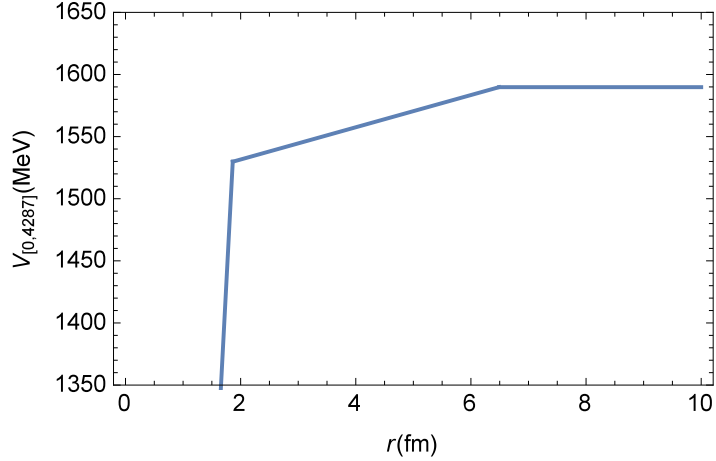}%
\caption{Representation of the $0^{-}(1^{--})$ $c\overline{c}$ static
potential $V_{\left[  0,m_{D\overline{D_{1}}}\right]  }(r)$ with
$m_{c}=1348.6$ MeV, $\sigma=850$ MeV/fm, $\zeta=100$ MeV.fm, $m_{D\overline
{D_{1}}}=4287$ MeV, $\Delta=60$ MeV and $s=12.97$ MeV/fm.}%
\label{Fig4}%
\end{center}
\end{figure}

In Fig. \ref{Fig4} we have plotted the potential $V_{\left[  0,m_{D\overline
{D_{1}}}\right]  }(r)$ for $\Delta=60$ MeV and $s\simeq13$ MeV/fm. As for the
threshold mass we use the value $m_{D\overline{D_{1}}}=4287$ MeV obtained from
the experimental masses of $D^{0}$ and $\overline{D_{1}}^{0}$ \cite{PDG18}.
For the remaining parameters we keep the formerly used values.%

The $0^{-}\left(  1^{--}\right)  $ low lying spectrum obtained from this
potential is shown in Table~\ref{Tab2}.

\begin{table}[ptb]%
\begin{tabular}
[c]{cccccc}%
$J^{PC}$ & $%
\begin{array}
[c]{c}%
\text{States}\\
nl_{\left[  0,\text{ }m_{D\overline{D_{1}}}\right]  }%
\end{array}
$ & $%
\begin{array}
[c]{c}%
m_{\left[  0,\text{ }m_{D\overline{D_{1}}}\right]  }\\
\text{MeV}%
\end{array}
$ & $%
\begin{array}
[c]{c}%
m_{PDG}\\
\text{MeV}%
\end{array}
$ & $%
\begin{array}
[c]{c}%
m_{Cor}\\
\text{MeV}%
\end{array}
$ & $%
\begin{array}
[c]{c}%
V_{C}\left(  r\right)  \text{ States }\\
nl
\end{array}
$\\\hline
$1^{--}$ &  &  &  &  & \\
& $1s_{\left[  0,m_{D\overline{D_{1}}}\right]  }$ & $3046.0$ & $3096.916\pm
0.011$ & $3046.0$ & $1s$\\
& $2s_{\left[  0,m_{D\overline{D_{1}}}\right]  }$ & $3632.1$ & $3686.09\pm
0.04$ & $3632.2$ & $2s$\\
& $1d_{\left[  0,m_{D\overline{D_{1}}}\right]  }$ & $3743.4$ & $3773.15\pm
0.33$ & $3743.5$ & $1d$\\
& $3s_{\left[  0,m_{D\overline{D_{1}}}\right]  }$ & $4061.0$ & $4039\pm1$ &
$4065.8$ & $3s$\\
& $2d_{\left[  0,m_{D\overline{D_{1}}}\right]  }$ & $4136.4$ & $4191\pm5$ &
$4142.8$ & $2d$\\
& $4s_{\left[  0,m_{D\overline{D_{1}}}\right]  }$ & $4261.5$ & $4230\pm8$ &  &
\\
& $3d_{\left[  0,m_{D\overline{D_{1}}}\right]  }$ & $4277.3$ &  &  &
\end{tabular}
\caption{Calculated $0^{-}\left(  1^{--}\right)  $ charmonium masses,
$m_{\left[  0,\text{ }m_{D\overline{D_{1}}}\right]  }$ from $V_{\left[
0,\text{ }m_{D\overline{D_{1}}}\right]  }(r)$ with $\sigma=850$ MeV/fm,
$\zeta=100$ MeV.fm, $m_{c}=1348.6$ MeV and $m_{D\overline{D_{1}}}=4287$ MeV.
The spectral notation $nl_{\left[  0,\text{ }m_{D\overline{D_{1}}}\right]  },$
where $n$ $\left(  l\right)  $ indicates the principal (orbital angular
momentum) quantum number, has been used for the states. Masses for
experimental resonances, $m_{PDG},$ have been taken from \cite{PDG18}. Masses
$m_{Cor}$ from $V_{C}(r)$, up to $m_{D\overline{D_{1}}}$, with the same values
for $\sigma,$ $\zeta$ and $m_{c}$ are also shown for comparison.}%
\label{Tab2}%
\end{table}

Notice that there is almost no difference between $V_{\left[  0,m_{D\overline
{D_{1}}}\right]  }(r)$ and $V_{C}(r)$ in the description of the (conventional)
sates below $4200$ MeV. On the contrary from this energy to threshold the use
of $V_{\left[  0,m_{D\overline{D_{1}}}\right]  }(r)$ gives rise to the
appearance of the $4s_{\left[  0,m_{D\overline{D_{1}}}\right]  }$ and
$3d_{\left[  0,m_{D\overline{D_{1}}}\right]  }$ states with no correspondence
at all with any conventional state from $V_{C}(r)$ (the $4s$ state has a mass
$m_{Cor}\left(  4s\right)  =4437$ MeV). This allows the accommodation of
$\psi(4260)$ as discussed in the next section.

\bigskip

For the sake of completeness it should be added that a non zero screening
energy interval potential, in line with lattice results, may also be used for
$0^{+}\left(  1^{++}\right)  $ states. However this does not give rise to any
significant difference with the zero screening energy interval approach used
in Section \ref{SII}. As a matter of fact for $\Delta_{1^{++}}=30$ MeV the
value of the slope can be chosen to get a completely equivalent description to
the one provided by the zero screening energy interval.

\section{$\psi(4260)$ \label{SIV}}

In Table~\ref{Tab2} the well established $\psi(4260)$ (different measurements
of its mass go from $4222$ MeV to $4284$ MeV; the quoted average mass in
\cite{PDG18} is $m_{\psi(4260)}=4230\pm8$ MeV) has been assigned to the
$4s_{\left[  0,m_{D\overline{D_{1}}}\right]  }$ state with a calculated mass
of $4261.5$ MeV although it is very probable that this state mixes with the
$3d_{\left[  0,m_{D\overline{D_{1}}}\right]  }$ one giving rise a mass closer
to the quoted experimental average. Under this assignment $\psi(4260)$ is an
unconventional state coming out from the string breaking effect due to the
$D\overline{D_{1}}$ threshold.

\begin{figure}
[ptb]
\begin{center}
\includegraphics[
height=2.4111in,
width=3.659in
]%
{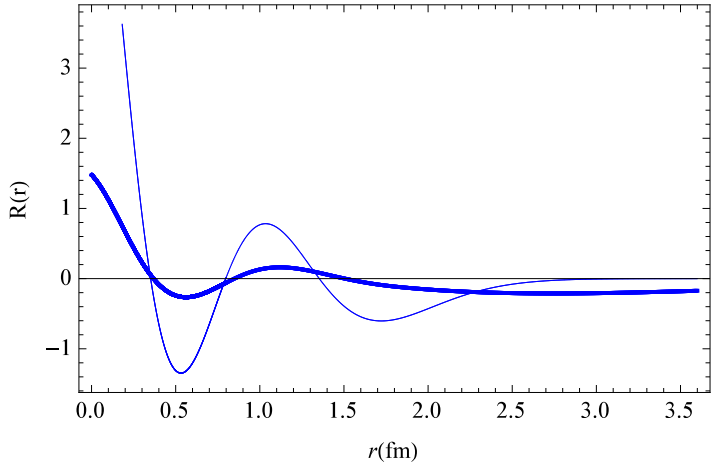}%
\caption{Radial wave functions R(\textit{r}) (in units $fm^{-\frac{3}{2}}$)
for the $1^{--}\left(  4s_{\left[  0,m_{D\overline{D_{1}}}\right]  }\right)  $
state (thick line) and the $1^{--}\left(  4s\right)  $ state (thin line).}%
\label{Fig5}%
\end{center}
\end{figure}

The role played by the $D\overline{D_{1}}\left(  2420\right)  $ configuration
has been previously emphasized by some authors, see for example \cite{Guo17},
\cite{ROS06}, \cite{DIN09}, \cite{WAN13}, \cite{GUO13}, \cite{QIN16} (and more
references therein). In our potential quark model the \textquotedblleft
molecular constituents\textquotedblright, $D\overline{D_{1}}\left(
2430\right)  $ and\ $D\overline{D_{1}}\left(  2420\right)  $ are embedded in
the quark-antiquark $4s_{\left[  0,m_{D\overline{D_{1}}}\right]  }$ radial
wave function, drawn in Fig. \ref{Fig5}, as reflected by the value of its root
mean square radius $\left\langle r^{2}\right\rangle ^{\frac{1}{2}}=3.75$ fm,
much larger than for wave functions from $V_{C}(r)$ (for instance,
$\left\langle r^{2}\right\rangle ^{\frac{1}{2}}=1.55$ fm for the $4s$ state
with a mass of $4437$ MeV). The non vanishing probability density at long
distances for the $4s_{\left[  0,m_{D\overline{D_{1}}}\right]  }$ state, say
the non vanishing probability for the heavy quark and antiquark to be far
apart clearly indicates that string breaking has taken place (as a related
consequence the probability density at the origin has been significantly reduced).%

\bigskip

One could argue that it is not a big deal to get the mass of a state through
the fixing of the free parameter $s.$ Nonetheless once we have the wave
function of $\psi(4260)$ we can calculate its decay properties and use their
comparison to data as a stringent test of our effective description. In this
regard let us remind that the discovery channel for $\psi(4260)$ was
$J/\psi\pi^{+}\pi^{-},$ that the conventionally dominant expected decay to
$D\overline{D}$ is suppressed, that the electromagnetic decay to $\chi
_{c1}\left(  3872\right)  \gamma$ is seen against the not seen decay to
$\chi_{c1}\left(  1p\right)  \gamma$ and that the following ratio has been
measured \cite{PDG18}
\begin{equation}
\left(  \frac{\Gamma_{\psi(4260)\rightarrow J/\psi\pi^{+}\pi^{-}}\Gamma
_{\psi(4260)\rightarrow e^{+}e^{-}}}{\Gamma_{\psi(4260)^{-}}}\right)
_{Exp}=9.2\pm1.0\text{ eV} \label{Expnumber}%
\end{equation}

\bigskip

It may be worth to mention that other screened potential models have been used
for the description of heavy quarkonia, see for example \cite{LiCh09},
\cite{Gon03}. These models use a general screened potential without connection
to any specific meson-meson threshold, yet generating a $4s$ state with a mass
about $4260$ MeV. In particular, in reference \cite{LiCh09} an analysis of
$\psi(4260)$ has been carried out (at the time of publication of reference
\cite{Gon03} the $\psi(4260)$ had not been discovered yet). As established by
the authors there are some considered difficulties, also shared by the other
screened potential models of the same kind, to assign the calculated $4s$
state to $\psi(4260).$ These difficulties have to do with the experimental
lack of coupling of $\psi(4260)$ to $e^{+}e^{-}$ and with the non observation
of the decay modes $D\overline{D},$ $D\overline{D}^{\ast}$ and $D^{\ast
}\overline{D}^{\ast}.$ Next we shall show that these difficulties are overcome
in our model signaling the need to include screening effects though a detailed
threshold consideration for a complete explanation of charmonium. Although we
shall rely on the particular choice $\left(  \Delta,s\right)  =\left(
60\text{ MeV, }13\text{ MeV/fm}\right)  $ we shall also give results for
$\left(  \Delta,s\right)  =\left(  120\text{ MeV, }64.2\text{ MeV/fm}\right)
$ and $\left(  \Delta,s\right)  =\left(  180\text{ MeV, }135\text{
MeV/fm}\right)  .$ This will allow us to establish the interval of variation
of $\Delta$ compatible with experimental observations.

\subsection{$\psi(4260)\rightarrow e^{+}e^{-}$}

For conventional $^{3}S_{1}$ bottomonium states below their corresponding $S-$
wave threshold the potential models we use, $V_{C}(r)$ and $V_{\left[
0,m_{D\overline{D_{1}}}\right]  }(r)$, reproduce quite approximately the
measured ratios of leptonic widths to $e^{+}e^{-}$. This ratios are calculated
as (see for example \cite{Eic08})
\begin{equation}
\frac{\Gamma_{i_{1}\rightarrow e^{+}e^{-}}}{\Gamma_{i_{2}\rightarrow
e^{+}e^{-}}}=\frac{\left\vert R_{i_{1}}\left(  0\right)  \right\vert ^{2}%
}{\left\vert R_{i_{2}}\left(  0\right)  \right\vert ^{2}}\frac{m_{i_{2}}^{2}%
}{m_{i_{1}}^{2}} \label{lwratio}%
\end{equation}
where $i_{1,2}$ stand for $^{3}S_{1}$ states, $R_{i_{1,2}}\left(  0\right)  $
for their radial wave functions at the origin and $m_{i_{1,2}}$ for their masses.

Regarding charmonium the calculated ratio $\frac{\Gamma_{2s_{\left[
0,m_{D\overline{D_{1}}}\right]  }\rightarrow e^{+}e^{-}}}{\Gamma_{1s_{\left[
0,m_{D\overline{D_{1}}}\right]  }\rightarrow e^{+}e^{-}}}=\frac{\Gamma
_{2s\rightarrow e^{+}e^{-}}}{\Gamma_{1s\rightarrow e^{+}e^{-}}}=0.5$ is a
$15\%$ off the experimental one $\left(  \frac{\Gamma_{\psi\left(  2s\right)
\rightarrow e^{+}e^{-}}}{\Gamma_{J/\psi\rightarrow e^{+}e^{-}}}\right)
_{Exp}=0.42\pm0.02.$

Then, by assuming a similar quality for the calculated ratios involving the
$4s_{\left[  0,m_{D\overline{D_{1}}}\right]  }$ state we can use
\begin{equation}
\frac{\Gamma_{\psi(4260)\rightarrow e^{+}e^{-}}}{\Gamma_{\psi\left(
2s\right)  \rightarrow e^{+}e^{-}}}\simeq\frac{\Gamma_{4s_{\left[
0,m_{D\overline{D_{1}}}\right]  }\rightarrow e^{+}e^{-}}}{\Gamma_{2s_{\left[
0,m_{D\overline{D_{1}}}\right]  }\rightarrow e^{+}e^{-}}}=\frac{\left\vert
R_{\psi(4260)}\left(  0\right)  \right\vert ^{2}}{\left\vert R_{\psi\left(
2s\right)  }\left(  0\right)  \right\vert ^{2}}\frac{m_{\psi\left(  2s\right)
}^{2}}{m_{\psi(4260)}^{2}}=2.4\times10^{-2} \label{e+e-ratio}%
\end{equation}
where $R_{\psi(4260)}\left(  0\right)  \simeq1.5$ fm$^{-\frac{3}{2}}$ and
$R_{\psi\left(  2s\right)  }\left(  0\right)  \simeq8.3$ fm$^{-\frac{3}{2}}$
from our model$,$ altogether with the experimental measurement $\left(
\Gamma_{\psi\left(  2s\right)  \rightarrow e^{+}e^{-}}\right)  _{Exp}%
=2.30\pm0.06$ KeV to predict an approximated leptonic decay width
\begin{equation}
\Gamma_{\psi(4260)\rightarrow e^{+}e^{-}}\simeq55.2\pm0.2\text{ eV}
\label{e+e-width}%
\end{equation}

Notice that this value is quite small as compared to $\left(  \Gamma
_{\psi\left(  2s\right)  \rightarrow e^{+}e^{-}}\right)  _{Exp}$ and other
values for conventional states. This is a direct consequence of the lack of
probability at the origin caused by screening expressed through the value of
the radial wave function at the origin.

Unfortunately the $\psi(4260)\rightarrow e^{+}e$ width has not been measured
separately for comparison. Instead\ we may use the experimentally known ratio%
\begin{equation}
\left(  \frac{\Gamma_{\psi(4260)\rightarrow J/\psi\pi^{+}\pi^{-}}\Gamma
_{\psi(4260)\rightarrow e^{+}e^{-}}}{\Gamma_{\psi(4260)}}\right)
_{Exp}=9.2\pm1.0\text{ eV} \label{IGS27}%
\end{equation}
to guess from (\ref{e+e-width}) the required branching ratio
\begin{equation}
\frac{\Gamma_{\psi(4260)\rightarrow J/\psi\pi^{+}\pi^{-}}}{\Gamma_{\psi
(4260)}}\simeq0.17\pm0.03 \label{IGS29}%
\end{equation}

Then from the total measured width
\begin{equation}
\left(  \Gamma_{\psi(4260)}\right)  _{Exp}=55\pm19\text{ MeV} \label{IGS28}%
\end{equation}
we get
\begin{equation}
\Gamma_{\psi(4260)\rightarrow J/\psi\pi^{+}\pi^{-}}\simeq9\pm5\text{ MeV}
\label{pipi}%
\end{equation}

\bigskip

It is worthwhile to point out that the leptonic width would be smaller than
the estimated value (\ref{e+e-width}) if $\psi(4260)$ contained also some
$3d_{\left[  0,m_{D\overline{D_{1}}}\right]  }$ probability. This would make
the branching ratio (\ref{IGS29}) and the decay width to $J/\psi\pi^{+}\pi
^{-}$ (\ref{pipi}) to increase their estimated values.

For the sake of consistency $\Gamma_{\psi(4260)\rightarrow J/\psi\pi^{+}%
\pi^{-}}$ should be reproduced from our quark model description. However, this
calculation involves the emission of two gluons through intermediate hybrid
states (see for instance \cite{Kua09}) that should be consistently obtained
within our quark model framework. This is a task out of the scope of the
present article.

Not withstanding this we should emphasize that a small value for $\Gamma
_{\psi(4260)\rightarrow e^{+}e^{-}}$ as the one we predict is a sine qua non
condition for $\psi(4260)\rightarrow J/\psi\pi^{+}\pi^{-}$ having a
significant branching ratio as required from being the discovery channel.
(Just for comparison, if we had used the $4s$ wave function to describe
$\psi(4260)$ the derived branching ratio would have been $0.005$.)
Furthermore, our predicted $\Gamma_{\psi(4260)\rightarrow e^{+}e^{-}}$is in
line with the experimental suppression of $S-$ wave $D\overline{D_{1}}$
production in $e^{+}e^{-}$ annihilation.

\bigskip

To study the dependence of these results on $\left(  \Delta,s\right)  $ we
have repeated the calculation for $\left(  \Delta,s\right)  =\left(  120\text{
MeV, }64.2\text{ MeV/fm}\right)  $ and $\left(  \Delta,s\right)  =\left(
180\text{ MeV, }135\text{ MeV/fm}\right)  .$ We get $\left(  \Gamma
_{\psi(4260)\rightarrow e^{+}e^{-}}\right)  _{\left(  120,64.2\right)  }%
\simeq230\pm6$ eV, $\left(  \Gamma_{\psi(4260)\rightarrow J/\psi\pi^{+}\pi
^{-}}\right)  _{\left(  120,64.2\right)  }\simeq2.2\pm1.2$ MeV and $\left(
\Gamma_{\psi(4260)\rightarrow e^{+}e^{-}}\right)  _{\left(  180,135\right)
}\simeq345\pm9$ eV, $\left(  \Gamma_{\psi(4260)\rightarrow J/\psi\pi^{+}%
\pi^{-}}\right)  _{\left(  180,135\right)  }\simeq1.5\pm0.8$ MeV. These values
are still compatible with data not permitting any discrimination among the
different $\Delta$ values. Incidentally, the predicted range of values for
$\Gamma_{\psi(4260)\rightarrow e^{+}e^{-}},$ $\left[  55\text{ eV},\text{
}345\text{ eV}\right]  ,$ is quite similar to the one expected from a
molecular model analysis \cite{Wan14}.

\subsection{E1 transitions}

For conventional bottomonium and charmonium states below their corresponding
$S-$ wave thresholds the potential models we use, $V_{C}(r)$ and $V_{\left[
0,m_{D\overline{D_{1}}}\right]  }(r)$, give correctly the order of magnitude
of the measured ratios of $^{3}S_{1}\leftrightarrow^{3}P_{1}$ dipole electric
transitions from the same initial state or to the same final state. More
accurate results are obtained if the experimental masses of the states are
used instead of the calculated ones.

The theoretical expressions for these ratios are:%

\begin{equation}
\frac{\Gamma_{E1}\left(  i\rightarrow f_{1}+\gamma\right)  }{\Gamma
_{E1}\left(  i\rightarrow f_{2}+\gamma\right)  }=\frac{w_{if_{1}}^{3}%
}{w_{if_{2}}^{3}}\frac{\left\vert D_{if_{1}}\right\vert ^{2}}{\left\vert
D_{if_{2}}\right\vert ^{2}} \label{iratio}%
\end{equation}
for the case in which the same initial state decays into two final $\left(
f_{1}\text{ and }f_{2}\right)  $ states with the same value of $J_{f}$ and
\begin{equation}
\frac{\Gamma_{E1}\left(  i_{1}\rightarrow f+\gamma\right)  }{\Gamma
_{E1}\left(  i_{2}\rightarrow f+\gamma\right)  }=\frac{w_{i_{1}f}^{3}%
}{w_{i_{2}f}^{3}}\frac{\left\vert D_{fi_{1}}\right\vert ^{2}}{\left\vert
D_{fi_{2}}\right\vert ^{2}} \label{fratio}%
\end{equation}
for the case in which two initial states $\left(  i_{1}\text{ and }%
i_{2}\right)  $ decay into the same final state.

$w_{if}$ is the photon energy and $D_{if}$ the electric dipole matrix element%
\begin{equation}
D_{if}=\int\limits_{0}^{\infty}drR_{i}(r)r^{2}\frac{3}{w_{if}}\left[
\frac{w_{if}r}{2}j_{0}\left(  \frac{w_{if}r}{2}\right)  -j_{1}\left(
\frac{w_{if}r}{2}\right)  \right]  R_{f}(r) \label{dif}%
\end{equation}
where $R_{i,f}(r)$ are the radial wave functions of the initial and final
mesons and $j_{0},$ $j_{1}$ stand for spherical Bessel functions.

By reasonably assuming the correct order of magnitude of the ratios when
transitions from $4s_{\left[  0,m_{D\overline{D_{1}}}\right]  }$ are involved
we predict (for $\psi(2s)\ $and $\chi_{c1}\left(  1p\right)  $ the
experimental masses are used; as for $\psi(4260)$ the calculated mass is taken
since we do not consider mixing with the $3d_{\left[  0,m_{D\overline{D_{1}}%
}\right]  }$ state)
\begin{equation}
\frac{\Gamma_{\psi(4260)\rightarrow\chi_{c1}\left(  3872\right)  \gamma}%
}{\Gamma_{\psi(4260)\rightarrow\chi_{c1}\left(  1p\right)  \gamma}}\simeq
\frac{\Gamma_{4s_{\left[  0,m_{D\overline{D_{1}}}\right]  \text{ }}%
\rightarrow\text{ }2p_{\left[  0,\text{ }m_{D\overline{D^{\ast}}}\right]
}\gamma}}{\Gamma_{4s_{\left[  0,m_{D\overline{D_{1}}}\right]  }\text{
}\rightarrow\text{ }1p_{\left[  0,\text{ }m_{D\overline{D^{\ast}}}\right]
}\gamma}}=107.8 \label{widthratio1}%
\end{equation}
and%

\begin{equation}
\frac{\Gamma_{\psi(4260)\rightarrow\chi_{c1}\left(  1p\right)  \gamma}}%
{\Gamma_{\psi(2s)\rightarrow\chi_{c1}\left(  1p\right)  \gamma}}\simeq
\frac{\Gamma_{4s_{\left[  0,m_{D\overline{D_{1}}}\right]  \text{ }}%
\rightarrow\text{ }1p_{\left[  0,\text{ }m_{D\overline{D^{\ast}}}\right]
}\gamma}}{\Gamma_{2s_{\left[  0,m_{D\overline{D_{1}}}\right]  }\text{
}\rightarrow\text{ }1p_{\left[  0,\text{ }m_{D\overline{D^{\ast}}}\right]
}\gamma}}=0.018 \label{widthratio2}%
\end{equation}

The first ratio (\ref{widthratio1}) provides an explanation for the decay
$\psi(4260)\rightarrow\chi_{c1}\left(  3872\right)  \gamma$ being seen against
the not seen decay $\psi(4260)\rightarrow\chi_{c1}\left(  1p\right)  \gamma.$
More quantitatively, we may use the second ratio (\ref{widthratio2}) to
predict from the experimental value $\left(  \Gamma_{\psi(2s)\rightarrow
\chi_{c1}\left(  1p\right)  \gamma}\right)  _{Exp}=29\pm1$ KeV a width
\[
\Gamma_{\psi(4260)\rightarrow\chi_{c1}\left(  1p\right)  \gamma}\simeq
0.506\pm0.017\text{ KeV}%
\]
Then from the first ratio we predict%
\[
\Gamma_{\psi(4260)\rightarrow\chi_{c1}\left(  3872\right)  \gamma}%
\simeq54.6\pm1.9\text{ KeV}%
\]

We should keep in mind though that according to our assumption above these
values of the widths should be considered as indicative of their order of
magnitude and not as accurate predictions.

\bigskip

As these radiative transitions are sensitive to the details of the wave
functions they can provide us, through its study from different $\left(
\Delta,s\right)  $ pairs, with some additional constraint on the $\Delta$
values. Actually the results we get $\left(  \Gamma_{\psi(4260)\rightarrow
\chi_{c1}\left(  1p\right)  \gamma}\right)  _{\left(  120,64.2\right)  }%
\simeq1.7\pm0.1$ keV, $\left(  \Gamma_{\psi(4260)\rightarrow\chi_{c1}\left(
3872\right)  \gamma}\right)  _{\left(  120,64.2\right)  }\simeq2.2\pm0.1$ keV
and $\left(  \Gamma_{\psi(4260)\rightarrow\chi_{c1}\left(  1p\right)  \gamma
}\right)  _{\left(  180,135\right)  }\simeq2.9\pm0.1$ keV, $\left(
\Gamma_{\psi(4260)\rightarrow\chi_{c1}\left(  3872\right)  \gamma}\right)
_{\left(  180,135\right)  }\simeq0.044\pm0.004$ keV, indicate that $\Delta$
should be smaller than $120$ MeV in order not to contradict the fact that the
decay $\psi(4260)\rightarrow\chi_{c1}\left(  3872\right)  \gamma$ is seen
whereas the $\psi(4260)\rightarrow\chi_{c1}\left(  1p\right)  \gamma$ decay is
not. Hence we may tentatively delimit the $\Delta$ interval to $\left[
60\text{ MeV, }120\text{ MeV}\right]  .$

\subsection{$\psi(4260)\rightarrow D\overline{D}$}

Other issue about $\psi(4260)$ has to do with the experimental suppression of
the $D\overline{D}$ decay mode despite the fact that $\psi(4260)$ is above the
$D\overline{D}$ threshold mass. In order to calculate this decay we shall rely
on the $^{3}P_{0}$ decay model \cite{LeY73,LeY77} where the physical mechanism
involved is related to the one we have used to take into account color
screening in the potential (a $q\overline{q}$ created in the hadronic vacuum
with $0^{++}$ quantum numbers combines with $\overline{c}c$ giving rise to
$D\overline{D}$). This model provides sensible results for the $D\overline{D}$
decay of the low lying conventional bottomonium and charmonium states with
mass above the $D\overline{D}$ threshold \cite{Ono81}.

Specifically the expression for the width is
\begin{equation}
\Gamma_{\psi(4260)\rightarrow D\overline{D}}=2\pi\frac{E_{D}E_{\overline{D}}%
}{m_{\psi(4260)}}k\left\vert A\right\vert ^{2} \label{widthformula}%
\end{equation}
where $E_{D}$ $\left(  =E_{\overline{D}}\right)  $ is the energy of the $D$
(or $\overline{D}$) meson given by
\begin{equation}
E_{D}=\sqrt{m_{D}^{2}+k^{2}}=E_{\overline{D}} \label{relenergy}%
\end{equation}
being $k$ the modulus of the three-momentum of $D$ (or $\overline{D}$) for
which we shall use the relativistic expression%
\begin{equation}
k=\frac{\sqrt{\left(  m_{\psi(4260)}^{2}-4m_{D}^{2}\right)  }}{2}
\label{threemom}%
\end{equation}
and $A$ stands for the decay amplitude given by%

\begin{equation}
\left\vert A\right\vert ^{2}\equiv\beta^{2}\left\vert \mathcal{M}\right\vert
^{2} \label{3p0amp}%
\end{equation}
where the constant $\beta$ specifies the strength of the pair creation, and
the expression for $\left\vert \mathcal{M}\right\vert ^{2}$ can be derived
from \cite{Ono81} in a straightforward manner (we use the same notation as in
this reference) as
\begin{equation}
\left\vert \mathcal{M}\right\vert ^{2}=\frac{1}{96}I\left(  +\right)  ^{2}
\label{MImenos}%
\end{equation}
where
\begin{equation}
I\left(  +\right)  ^{2}=\left\vert
\begin{array}
[c]{c}%
\frac{1}{\hbar^{\frac{9}{2}}}\int_{0}^{\infty}r_{X}^{2}dr_{X}\psi_{X}\left(
r_{X}\right)  \int p^{2}dp\widetilde{u}_{D}\left(  p\right)  \widetilde
{u}_{\overline{D}}\left(  p\right) \\
\left[  pj_{1}\left(  \frac{pr_{X}}{\hbar}\right)  j_{1}\left(  \frac{m_{c}%
}{\left(  m_{c}+m_{q}\right)  }\frac{kr_{X}}{\hbar}\right)  +\frac{m_{q}%
}{\left(  m_{c}+m_{q}\right)  }kj_{0}\left(  \frac{pr_{X}}{\hbar}\right)
j_{0}\left(  \frac{m_{c}}{\left(  m_{c}+m_{q}\right)  }\frac{kr_{X}}{\hbar
}\right)  \right]
\end{array}
\right\vert ^{2} \label{msquare}%
\end{equation}
$m_{q}=340$ MeV is the mass of the light quark, $\psi_{X}$ denotes the radial
wave function of $\psi(4260)$ in configuration space and $\widetilde{u}%
_{D}\left(  p\right)  $ stands for radial wave function of $D$ in momentum
space
\begin{equation}
\widetilde{u}_{D}\left(  p\right)  \equiv\sqrt{\frac{2}{\pi}}\int_{0}^{\infty
}r_{D}^{2}dr_{D}\psi_{D}\left(  r_{D}\right)  j_{0}\left(  \frac{pr_{D}}%
{\hbar}\right)  \label{momwfD}%
\end{equation}
calculated from $\psi_{D},$ the radial wave function of $D$ in configuration space.

In order to simplify the calculation we shall approach as usual $\psi
_{D}\left(  r_{D}\right)  $ by a gaussian (the same expression for
$\psi_{\overline{D}}\left(  r_{\overline{D}}\right)  $)%
\begin{equation}
\psi_{D}\left(  r_{D}\right)  =\frac{2}{\pi^{\frac{1}{4}}R_{D}^{\frac{3}{2}}%
}e^{-\frac{r_{D}^{2}}{2R_{D}^{2}}} \label{rwfD}%
\end{equation}
$R_{D}$ can be fixed either variationally or by requiring it to be equal to
the root mean square (rms) radius obtained from the description of
(conventional) $D$ with $V_{C}(r)$ and a light quark mass of about $340$ MeV
(this implies a change of the value of the coulomb strength $\zeta$ to get the
spectral mass). By using the rms radius procedure we get $R_{D}=0.54$ fm. Then
the use of the gaussian instead of the wave function from $V_{C}(r)$ hardly
makes any difference.

\bigskip

We may avoid the dependence on the constant $\beta$ by taking the ratio with
some other $D\overline{D}$ decay process. Furthermore if the width for this
other process has been measured then we can give a prediction for
$\Gamma_{\psi(4260)\rightarrow D\overline{D}}$ by assuming that the calculated
ratio approximates the experimental one. These conditions may be satisfied by
choosing the process $\psi(3770)\rightarrow D\overline{D}.$ (Notice that
$\psi(3770)$ has been assigned to the $1d_{\left[  0,m_{D\overline{D_{1}}%
}\right]  }$ state in Table~\ref{Tab2}.)

The $\psi(3770)\rightarrow D\overline{D}$ width is given by%
\[
\Gamma_{\psi(3770)\rightarrow D\overline{D}}=2\pi\frac{E_{D}^{\prime
}E_{\overline{D}}^{\prime}}{m_{\psi(3770)}}k^{\prime}\left\vert A^{\prime
}\right\vert ^{2}%
\]
with%
\[
k^{\prime}=\frac{\sqrt{\left(  m_{\psi(3770)}^{2}-4m_{D}^{2}\right)  }}{2}%
\]
and $\left\vert A^{\prime}\right\vert ^{2}\equiv\beta^{2}\left\vert
\mathcal{M}^{\prime}\right\vert ^{2}$ with
\[
\left\vert \mathcal{M}^{\prime}\right\vert ^{2}=\frac{1}{48}I\left(  -\right)
^{2}%
\]
and%
\[
I\left(  -\right)  ^{2}=\left\vert
\begin{array}
[c]{c}%
\frac{1}{\hbar^{\frac{9}{2}}}\int_{0}^{\infty}r_{X}^{2}dr_{X}\psi_{X}^{\prime
}\left(  r_{X}\right)  \int p^{2}dp\widetilde{u}_{D}\left(  p\right)
\widetilde{u}_{\overline{D}}\left(  p\right) \\
\left[  -pj_{1}\left(  \frac{pr_{X}}{\hbar}\right)  j_{1}\left(  \frac{m_{c}%
}{\left(  m_{c}+m_{q}\right)  }\frac{k^{\prime}r_{X}}{\hbar}\right)
+\frac{m_{q}}{\left(  m_{c}+m_{q}\right)  }k^{\prime}j_{0}\left(  \frac
{pr_{X}}{\hbar}\right)  j_{2}\left(  \frac{m_{c}}{\left(  m_{c}+m_{q}\right)
}\frac{k^{\prime}r_{X}}{\hbar}\right)  \right]
\end{array}
\right\vert ^{2}%
\]
where $\psi_{X}^{\prime}$ denotes the radial wave function of $\psi(3770)$ in
configuration space.

\bigskip

By making use of these expressions we get (the experimental mass for
$\psi(3770)$ has been used)
\[
\frac{\Gamma_{\psi(4260)\rightarrow D\overline{D}}}{\Gamma_{\psi
(3770)\rightarrow D\overline{D}}}=7\times10^{-3}%
\]
that explains the $D\overline{D}$ decay suppression for $\psi(4260)$ as
compared to the conventional $\psi(3770)$ state. Quantitatively, using this
ratio and the measured values
\[
\left(  \Gamma_{\psi(3770)\rightarrow D\overline{D}}\right)  _{Exp}%
=25.6\pm0.8\text{ MeV}%
\]
and
\[
\left(  \Gamma_{\psi(4260)}\right)  _{Exp}=55\pm19\text{ MeV}%
\]
we predict%
\[
\Gamma_{\psi(4260)\rightarrow D\overline{D}}\simeq0.18\pm0.01\text{ MeV}%
\]
and
\[
\frac{\Gamma_{\psi(4260)\rightarrow D\overline{D}}}{\left(  \Gamma
_{\psi(4260)}\right)  _{Exp}}\simeq\left(  3\pm2\right)  \times10^{-3}%
\]
Regarding the dependence on the constrained $\left(  \Delta,s\right)  $
interval we get $\left(  \Gamma_{\psi(4260)\rightarrow D\overline{D}}\right)
_{\left(  120,64.2\right)  }\simeq1.7\pm0.1$ MeV, which is still suppressed
with respect to $\Gamma_{\psi(3770)\rightarrow D\overline{D}}$ by a factor
$15$. Hence we may expect the experimental suppression factor to be comprised
in the interval $\left[  143,15\right]  $. To go beyond in the determination
of this factor one should make use of it to calculate how the cross section
$\sigma\left(  e^{+}e^{-}\rightarrow D\overline{D}\right)  $ differs at centre
of mass energies of $3770$ MeV and $4260$ MeV$.$ Then, through a comparison to
the measured values of $R=\frac{\sigma_{tot}\left(  e^{+}e^{-}\rightarrow
hadrons\right)  }{\sigma_{QED}\left(  e^{+}e^{-}\rightarrow\mu^{+}\mu
^{-}\right)  }$ at these energies \cite{Bai02}, a more precise value of the
factor might be estimated. This is a quite interesting program but clearly out
of the scope of this article.

\section{Summary\label{SV}}

Starting from lattice results for the energy of two static color sources ($Q$
and $\overline{Q}$) when mixing of the $Q\overline{Q}$ configuration with an
open flavor meson-meson one is taken into account the form of a
Born-Oppenheimer quark-antiquark static potential can be prescribed. This
potential contains implicitly the effect of color screening due to the
presence of light $q\overline{q}$ pairs that combine with $Q\overline{Q}$
giving rise to meson-meson components.

A simplified prescription corresponding to consider that screening takes place
just at the meson-meson threshold energy, previously used for the description
of $0^{+}\left(  J^{++}\right)  $ charmonium $\left(  J=0,1,2\right)  ,$ has
been refined by the introduction of a non zero screening energy interval to
deal with $0^{-}\left(  1^{--}\right)  $ states below their first $S-$ wave
meson-meson threshold. The spectrum from the resulting potential contains
conventional like states as well as unconventional ones. This allows for the
theoretical accommodation of the experimentally well established resonance
$\psi(4260)$ through its assignment to a calculated sate$.$ To check the
viability of such an assignment we have calculated $e^{+}e^{-}$, $E1$ and
$D\overline{D}$ decay widths. Our results show full compatibility with
existing data although more refined measurements would be needed for a more
detailed comparison. Meanwhile we may tentatively conclude that $\psi(4260)$
may be described as an unconventional state coming out from the string
breaking effect due to $D\overline{D_{1}}$ meson-meson components.

\bigskip

\bigskip

This work has been supported by \foreignlanguage{spanish}{Ministerio de Economía y Competitividad} of Spain (MINECO) and EU Feder grant FPA2016-77177-C2-1-P and by SEV-2014-0398. R. B. acknowledges the \foreignlanguage{spanish}{Ministerio de Ciencia, Innovación y Universidades} of Spain for a FPI fellowship.

\end{document}